\documentclass[conference]{IEEEtran}
\usepackage{color,graphicx,amsmath,amssymb,stmaryrd,dsfont}

\DeclareMathAlphabet{\eurm}{U}{eur}{m}{n}
\DeclareMathAlphabet{\mathbsf}{OT1}{cmss}{bx}{n}
\DeclareMathAlphabet{\mathssf}{OT1}{cmss}{m}{sl}
\DeclareMathAlphabet{\mathcsf}{OT1}{cmss}{sbc}{n}

\newcommand{\randomvalue}[1]{{\uppercase{#1}}}

\DeclareSymbolFont{bsfletters}{OT1}{cmss}{bx}{n}  
\DeclareSymbolFont{ssfletters}{OT1}{cmss}{m}{n}
\DeclareMathSymbol{\bsfGamma}{0}{bsfletters}{'000}
\DeclareMathSymbol{\ssfGamma}{0}{ssfletters}{'000}
\DeclareMathSymbol{\bsfDelta}{0}{bsfletters}{'001}
\DeclareMathSymbol{\ssfDelta}{0}{ssfletters}{'001}
\DeclareMathSymbol{\bsfTheta}{0}{bsfletters}{'002}
\DeclareMathSymbol{\ssfTheta}{0}{ssfletters}{'002}
\DeclareMathSymbol{\bsfLambda}{0}{bsfletters}{'003}
\DeclareMathSymbol{\ssfLambda}{0}{ssfletters}{'003}
\DeclareMathSymbol{\bsfXi}{0}{bsfletters}{'004}
\DeclareMathSymbol{\ssfXi}{0}{ssfletters}{'004}
\DeclareMathSymbol{\bsfPi}{0}{bsfletters}{'005}
\DeclareMathSymbol{\ssfPi}{0}{ssfletters}{'005}
\DeclareMathSymbol{\bsfSigma}{0}{bsfletters}{'006}
\DeclareMathSymbol{\ssfSigma}{0}{ssfletters}{'006}
\DeclareMathSymbol{\bsfUpsilon}{0}{bsfletters}{'007}
\DeclareMathSymbol{\ssfUpsilon}{0}{ssfletters}{'007}
\DeclareMathSymbol{\bsfPhi}{0}{bsfletters}{'010}
\DeclareMathSymbol{\ssfPhi}{0}{ssfletters}{'010}
\DeclareMathSymbol{\bsfPsi}{0}{bsfletters}{'011}
\DeclareMathSymbol{\ssfPsi}{0}{ssfletters}{'011}
\DeclareMathSymbol{\bsfOmega}{0}{bsfletters}{'012}
\DeclareMathSymbol{\ssfOmega}{0}{ssfletters}{'012}

\newcommand{\rvM}{{\randomvalue{M}}}	
\newcommand{\rvQ}{{\randomvalue{Q}}}	
\newcommand{\rvU}{{\randomvalue{U}}}	
\newcommand{\rvV}{{\randomvalue{V}}}	
\newcommand{\rvW}{{\randomvalue{W}}}	
\newcommand{\rvX}{{\randomvalue{X}}}  	
\newcommand{\rvZ}{{\randomvalue{Z}}}	

\newcommand{\calC}{{\mathcal{C}}}

\newcommand{\calM}{{\mathcal{M}}}

\newcommand{\calO}{{\mathcal{O}}}

\newcommand{\calR}{{\mathcal{R}}}

\newcommand{\calS}{{\mathcal{S}}}
\newcommand{\calU}{{\mathcal{U}}}
\newcommand{\calV}{{\mathcal{V}}}
\newcommand{\calX}{{\mathcal{X}}}

\newcommand{\calW}{{\mathcal{W}}}

\newcommand{\E}[2][]{{\mathbb{E}_{#1}}{\left(#2\right)}}       
       
\newcommand{\V}[1]{{{\mathbb{V}}\!\left(#1\right)}}
\newcommand{\avgI}[1]{{{\mathbb{I}}\!\left(#1\right)}}

\newcommand{\avgH}[1]{{\mathbb{H}}\!\left(#1\right)}

\newcommand{\sts}[3]{T_{#1}^{#2}\!(#3)}

\newcommand{\card}[1]{\ensuremath{\left|{#1}\right|}}           
\newcommand{\abs}[1]{\ensuremath{\left|#1\right|}}              
\newcommand{\eqdef}{\ensuremath{\triangleq}}                    
\newcommand{\intseq}[2]{\ensuremath{\llbracket{#1},{#2}\rrbracket}}  

\newtheorem{theorem}{Theorem}

\newtheorem{definition}{Definition}

\newtheorem{corollary}{Corollary}

\renewcommand{\leq}{\leqslant}
\renewcommand{\geq}{\geqslant}

\newcommand{\jkl}[1]{\textcolor{black}{#1}}

\graphicspath{{graphics/}}

\begin{document}

\title{Strong Coordination over a Line Network}

\author{\IEEEauthorblockN{Matthieu R. Bloch}
  \IEEEauthorblockA{School of Electrical and Computer Engineering\\
Georgia Institute of Technology\\
Atlanta, Georgia 30332--0250\\
Email: matthieu.bloch@ece.gatech.edu}
\and
\IEEEauthorblockN{J{\"o}rg Kliewer}
\IEEEauthorblockA{Klipsch School of Electrical and Computer Engineering\\
New Mexico State University\\
Las Cruces, New Mexico 88003-8001\\
Email: jkliewer@nmsu.edu}
}

\maketitle

\begin{abstract}
We study the problem of strong coordination in a three-terminal line
network, in which agents use common randomness and communicate over a line
network to ensure that their actions follow a prescribed behavior, modeled
by a target joint distribution of actions. We provide inner and outer bounds
to the coordination capacity region, and show that these bounds are
partially optimal. We leverage this characterization to develop insight into
the interplay between communication and coordination. Specifically, we show
that common randomness helps to achieve optimal communication rates between agents, and that matching the network topology to the behavior structure may reduce inter-agent communication rates.
\end{abstract}

\section{Introduction}
\label{sec:introduction}

One fundamental problem in decentralized networked systems is to coordinate
activities of different agents so that they reach a state of agreement. In this paper, we measure coordination by the ability to achieve a prescribed joint probability distribution of actions at all agents in the network. The information-theoretic limits of such a coordination have been partly characterized~\cite{CPC10} using the notions of \emph{empirical coordination}, which only requires the normalized histogram of induced joint actions to approach a desired target distribution, and \emph{strong coordination}, where the sequence of induced joint actions must be statistically indistinguishable from the target distribution. If the actions to coordinate are dependent, it has been shown that the communication rate among agents can be significantly reduced compared to a basic approach of communicating explicit messages describing the actions. Further study of the fundamental limits of networked coordination could help develop insight into the interplay between communication and coordination, which could in turn guide the design of many applications, for example in distributed control or multi-agent based exploration and surveillance. 

The information-theoretic limits of empirical coordination for small and large networks have been the subject of several investigations. For instance, \cite{KS07} studies the rate required to reconstruct the empirical distribution of a source at the output of a communication channel rather than reconstructing the source itself. Further,~\cite{BSST02,Cuff2008} analyze the rate of noise-free communication required to mimic a noisy memoryless communication channel. The work in~\cite{AB07} considers a distributed multi-agent control problem, in which each agent generates actions based on its own observations of a source of randomness. Recently, \cite{BTS12,Bloch2012a} have proposed coordination schemes based on polar codes achieving empirical coordination and strong coordination for specific distributions of actions. The generation of dependent random variables in networks under a strong coordination constraint is considered in~\cite{GA11,YGA12,Haddadpour2012}, by considering bidirectional transmissions in several rounds; however, these works only address the coordination of two nodes. 

In this paper, we attempt to develop further insight into the relation between network communication topology and coordination by extending the work in~\cite{CPC10} about point-to-point strong coordination to a three-terminal line network. We provide inner and outer bounds to the coordination capacity region and characterize the optimal communication rates between agents. We also analyze the impact of the underlying network topology on the minimization of the communication requirements for coordination.

\section{Problem Statement and Main Results}

\subsection{Motivating scenario}
\label{sec:motiv-prel}
\begin{figure}[t]
  \centerline{\includegraphics[scale=0.5]{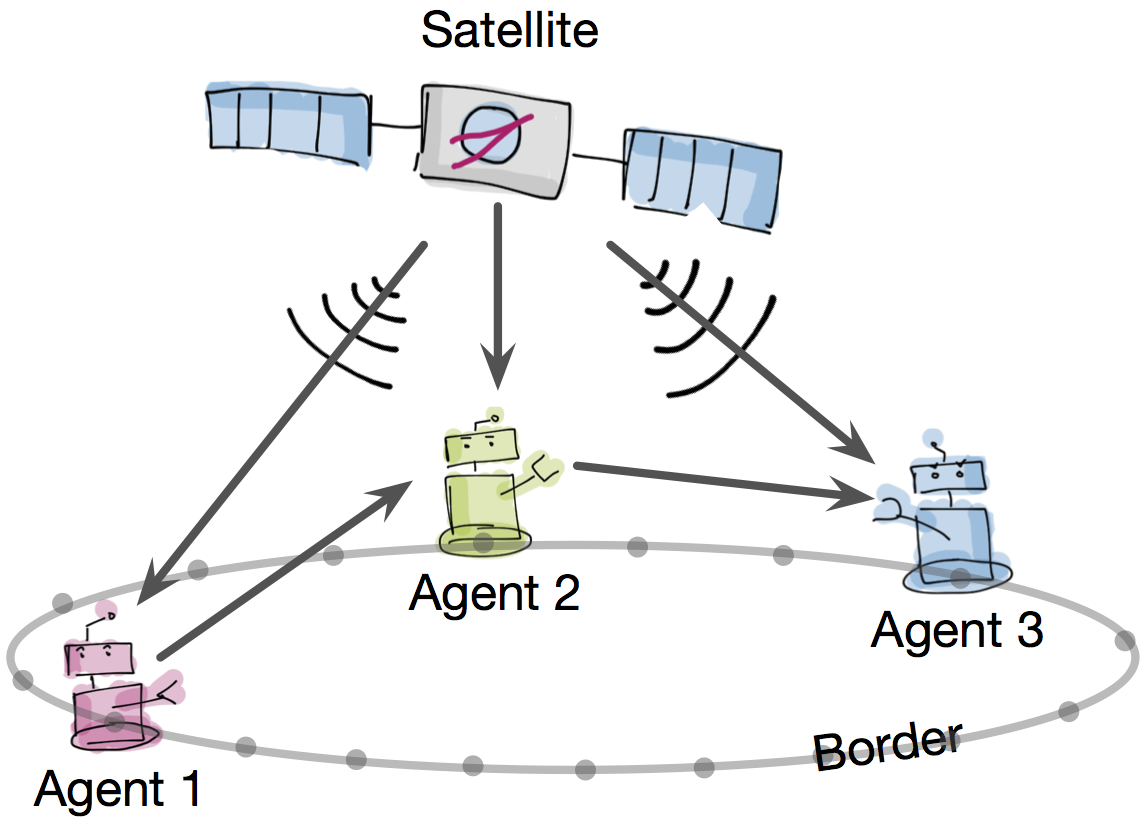}}
  \caption{Coordination of three agents in a perimeter defense scenario.}
  \label{fig:robot_example}
\end{figure}

As a motivation, consider the perimeter defense scenario illustrated in  Fig.~\ref{fig:robot_example}, in which three agents patrol a
border to avoid intrusions. 
Each agent is able to take sequences of binary actions (``move left'',
``move right'') described by the random variables $X_i\sim\text{Bern}(p_i)$,
and the objective is to communicate to achieve a desired behavior, captured
by a prescribed joint distribution $q_{X_1X_2X_3}$ of the actions. The
agents also have access to common randomness, which is pictorially
illustrated by an overhead satellite. \jkl{In this scenario, it is possible to
deploy several communication network topologies to coordinate the actions of
the agents, and we are interested in determining the optimal topology that
minimizes the inter-agent communication rate.} As a first step towards this
goal, we study the problem of coordination along a \emph{line network} with unidirectional communication,
in which Agent~1 can only communicate with Agent~2, and Agent~2 can only
communicate with Agent~3, possibly assisted by common randomness.

\subsection{Problem setting and main result}
\label{sec:problem-setting}

\label{sec:problem-statement}
\begin{figure}
  \centering
  \scalebox{0.55}{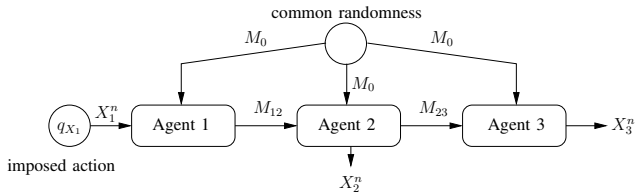}
  \caption{Coordination over a line network.}
  \label{fig:coordination_setting}
\end{figure}
Formally, we consider the setting illustrated in Fig.~\ref{fig:coordination_setting}, in which three agents wish to coordinate their actions. The actions taken by Agent $i\in\{1,2,3\}$ is described by a sequence of discrete actions $x_i^n\in\calX_i^n$, and the behavior is captured by the joint probability distribution of the actions. The network has a line structure in the sense that there are only two communication links:
\begin{itemize}
\item a noiseless link between Agent 1 and Agent 2, over which Agent 1 transmit messages in the set $\calM_{12}\eqdef\intseq{1}{2^{nR_{12}}}$;
\item a noiseless link between Agent 2 and Agent 3, over which Agent 2 transmit messages in the set $\calM_{23}\eqdef\intseq{1}{2^{nR_{23}}}$.
\end{itemize}
In addition, we assume that all agents have access to a common source of randomness, which produces uniform random numbers in the set $\calM_0\eqdef\intseq{1}{2^{nR_0}}$, and that Agent 1 independently determines his own sequence of actions according to the prescribed distribution $q_{X_1^n}$; the objective is then to control the actions of Agent 2 and Agent 3 by means of a line coordination code defined as follows.
\begin{definition}
  \label{def:line_coordination}
  A $(2^{nR_0},2^{nR_{12}},2^{nR_{23}},n)$ line coordination code consists of:
  \begin{itemize}
  \item an encoder $f_{12}:\calM_0\times\calX_1^n\rightarrow\calM_{12}$ to send messages from Agent 1 to Agent 2;
  \item an actuator $g_2:\calM_0\times\calM_{12}\rightarrow\calX_2^n$ to generate the actions of Agent 2;
  \item an encoder $f_{23}:\calM_0\times\calM_{12}\rightarrow\calM_{23}$ to send messages from Agent 2 to Agent 3;
  \item an actuator $g_3:\calM_0\times\calM_{23}\rightarrow\calX_3^n$ to generate the actions of Agent 3.
  \end{itemize}
\end{definition}
The communication and processing at each agent induces joint actions characterized by the joint distribution of actions $p(x_1^n,x_2^n,x_3^n)$. The goal is to design a code so that $p(x_1^n,x_2^n,x_3^n)$ is arbitrarily close to the target joint distribution $q(x_1^n,x_2^n,x_3^x)\eqdef\prod_{i=1}^nq_{\rvX_1\rvX_2\rvX_3}(x_{1,i},x_{2,i},x_{3,i})$. Formally, a rate triplet $(R_0,R_{12}, R_{23})$ is achievable if there exists a sequence of $(2^{nR_0},2^{nR_{12}},2^{nR_{23}},n)$ line coordination codes with increasing length $n$ such that
\begin{align*}
  \lim_{n\rightarrow\infty}\V{p_{X_1^nX_2^nX_3^n},q_{X_1^n,X_2^n,X_3^n}}=0,
\end{align*}
where $\mathbb{V}$ denotes the, \jkl{$L_1$-distance
between two distributions}. The set of all achievable rate triplets is called the coordination capacity region $\calC(q_{\rvX_1\rvX_2\rvX_3})$, and the central result of this paper is a partial characterization of $\calC(q_{\rvX_1\rvX_2\rvX_3})$.

As a baseline, let us start by considering a simple line coordination code, in which Agent~1 takes a sequence of actions $x_1^n$ drawn according to $q_{X_1^n}$ and sends an \emph{explicit} description of $x_1^n$
to Agent~2; this requires a rate of $R_{12}\geq H(X_1)$ bits/action. Agent~2 then takes a sequence of actions $x_2^n$ drawn according to
$q_{X_2^n|X_1^n=x_1^n}$ and sends an explicit description of both $x_1^n$ and $x_2^n$ to Agent 3; this requires a rate of $R_{23}\geq H(X_1X_2)$ bits/action. Agent 3 finally takes a sequence of actions $x_3^n$ generated according to $q_{X_3^n|X_2^n=x_2^n,X_1^n=x_1^n}$. Note that this code does not exploit common randomness. Consequently, the rates achievable by this
code are
\begin{equation}
    \left\{
          (R_0,R_{12},R_{23}):
          \begin{array}{l}
            R_0\geq 0\\
            R_{12}\geq H(X_1)\\
            R_{23} \geq H(X_1X_2)
          \end{array}
        \right\}.
        \label{eq:benchmark_rates}
\end{equation}
We show next that a much improved rate region can be achieved by exploiting common randomness.

Our result is expressed in terms of the following short-hand notation. Let $\rvU\in\calU$, $\rvV\in\calV$, and $\rvV\in\calW$ be auxiliary random variables. We define  the sets $\calS_{\text{in}}$ and $\calS_{\text{out}}$ of joint distributions on $\calU\times\calV\times\calW\times\calX_1\times\calX_2\times\calX_3$ as
\begin{align}
\calS_{\text{out}}&\eqdef\left\{p_{\rvU\rvV\rvW\rvX_1\rvX_2\rvX_3}: 
  \begin{array}{l}
    p_{\rvX_1\rvX_2\rvX_3}=q_{\rvX_1\rvX_2\rvX_3}\\
    \rvX_1\rightarrow\rvU\rvW\rightarrow\rvV\rvX_2\rvX_3\\
    \rvX_1\rvX_2\rvU\rightarrow\rvV\rvW\rightarrow\rvX_3\\
  \end{array}
\right\},\\
  \calS_{\text{in}}&\eqdef\left\{p_{\rvU\rvV\rvW\rvX_1\rvX_2\rvX_3}:
  \begin{array}{l}
    p_{\rvU\rvV\rvW\rvX_1\rvX_2\rvX_3}\in\calS_{\text{out}}\\
    \rvU\rightarrow\rvW\rightarrow\rvV
  \end{array}
\right\}.
\end{align}
For a fixed distribution $p_{\rvU\rvV\rvW\rvX_1\rvX_2\rvX_3}$, we also define the rate region
\begin{multline}
  \calR(p_{\rvU\rvV\rvW\rvX_1\rvX_2\rvX_3})\eqdef\\
  \left\{
    \begin{array}{l}
      (R_0,R_{12},R_{23}):\\
      R_0+R_{12}+R_{23}\geq \avgI{\rvU\rvV\rvW;\rvX_1\rvX_2\rvX_3}\\
      R_0+R_{12} \geq\avgI{\rvU\rvW;\rvX_1\rvX_2\rvX_3}\\
      R_0+R_{23}\geq \avgI{\rvV\rvW;\rvX_1\rvX_2\rvX_3}\\
      R_0\geq \avgI{\rvW;\rvX_2\rvX_2\rvX_3}\\
      R_{12}+R_{23}\geq \avgI{\rvU\rvV\rvW;\rvX_1}\\
      R_{12}\geq \avgI{\rvU\rvW;\rvX_1}\\
      R_{23}\geq \avgI{\rvV\rvW;\rvX_1}
    \end{array}
\right\}.\label{eq:region}
\end{multline}

\begin{theorem}
  \label{th:main_result}
  The coordination capacity region $\calC(q_{\rvX_1\rvX_2\rvX_3})$ of the line network satisfies
  \begin{align*}
    \bigcup_{p\in\calS_{\text{in}}}  \calR(p)\subseteq \calC(q_{\rvX_1\rvX_2\rvX_3}) \subseteq     \bigcup_{p\in\calS_{\text{out}}}  \calR(p).
\end{align*}
\end{theorem}
\begin{IEEEproof}
  For clarity, the proofs are relegated to Section~\ref{sec:achievability-proof} and Section~\ref{sec:converse-proof}, and we do not provide the cardinality bounds on the auxiliary random variables.
\end{IEEEproof}
Note that the inner and outer bounds for $\calC(q_{\rvX_1\rvX_2\rvX_3})$ may not match because $\calS_{\text{in}}$ is a strict subset of $\calS_{\text{out}}$, in general. The constraint $\rvU\rightarrow\rvW\rightarrow\rvV$ prevents us from choosing $\rvW$ independently of $\rvX_1\rvX_2\rvX_3$, so that $R_0\geq \avgI{\rvW;\rvX_2\rvX_2\rvX_3}>0$ in general, \jkl{as $\rvX_1\rvX_2\rvX_3$ depend on $\rvU\rvV$ (see Section~\ref{sec:achievability-proof})}. Consequently, the proposed coding scheme does not specialize to the baseline scheme whose rates are given in Eq.~\eqref{eq:benchmark_rates}.

\subsection{Applications}
\label{sec:applications}

The multiple auxiliary random variables involved in Theorem~\ref{th:main_result} make it rather difficult to parse the bounds obtained for $\calC(q_{\rvX_1\rvX_2\rvX_3})$. To obtain additional insight, we specialize Theorem~\ref{th:main_result} and characterize the region ${\calC}^*(q_{\rvX_1\rvX_2\rvX_3})$, defined as the projection of $\calC(q_{\rvX_1\rvX_2\rvX_3})$ onto the plane $R_0=0$. In other words, the region ${\calC}^*(q_{\rvX_1\rvX_2\rvX_3})$ characterizes the optimal communication rates between the agents for coordination, assuming the rate of common randomness can be chosen arbitrarily.

\begin{corollary}
  \label{cor:projection_region}
  The coordination capacity region $\calC^*(q_{\rvX_1\rvX_2\rvX_3})$ is characterized by
  \begin{align*}
    \calC^*(q_{\rvX_1\rvX_2\rvX_3}) = \left\{
        (R_{12},R_{23}):      \begin{array}{l}
        R_{12}\geq\avgI{\rvX_2\rvX_3;\rvX_1}\\
        R_{23}\geq\avgI{\rvX_3;\rvX_1}
      \end{array}
\right\}.
  \end{align*}
\end{corollary}
\begin{IEEEproof}
  We first simplify the outer bound of Theorem~\ref{th:main_result} by noting the following. Because of the Markov chains $\rvX_1-\rvU\rvW-\rvV\rvX_2\rvX_3$ and $\rvX_1\rvX_2\rvU-\rvV\rvW-\rvX_3$, the constraint on the sum rate $R_{12}+R_{23}$ in Eq.~\eqref{eq:region} is ineffective and subsumed by the constraint on $R_{12}$. In addition, the data processing inequality ensures that
  \begin{align*}
    \avgI{\rvU\rvW;\rvX_1}\geq \avgI{\rvX_2\rvX_3;\rvX_1}\text{ and }\avgI{\rvV\rvW;\rvX_1}\geq \avgI{\rvX_3;\rvX_1}.
\end{align*}

To show that the region is achievable, note that the choice $\rvU\eqdef\rvX_2$, $\rvW\eqdef\rvX_3$, $\rvV\eqdef {\rvX_3}$ satisfies the constraints of the set $\calS_{\text{in}}$, so that it may be substituted into the inner bound. This choice directly yields the desired result.
\end{IEEEproof}

We now develop several insights regarding the interplay between coordination and communication topology by leveraging the simple expression of Corollary~\ref{cor:projection_region}, which does not involve any auxiliary random variables.\smallskip{}

\textbf{Insight 1: Common randomness helps.} Comparing the result of
Corollary~\ref{cor:projection_region} with the baseline performance in
Eq.~\eqref{eq:benchmark_rates}, we observe that
$\avgI{\rvX_2\rvX_3;\rvX_1}\leq \avgH{\rvX_1}$ and $\avgI{\rvX_3;\rvX_1}\leq
\avgH{\rvX_1\rvX_2}$, so that the baseline scheme is suboptimal, in general.
\smallskip{}

\begin{figure}[t]
  \centering
    \scalebox{0.55}{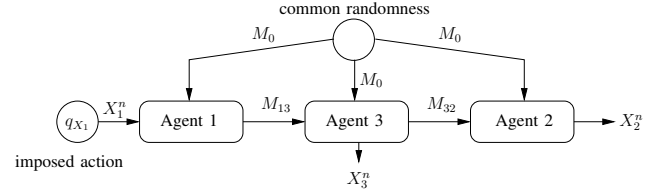}
  \caption{Modified line network topology.}
  \label{fig:modified_line}
\end{figure}

\textbf{Insight 2: The communication topology should match the coordination structure.} Assume now that the desired behavior $q_{\rvX_1\rvX_2\rvX_3}$ is such that $\rvX_1-\rvX_2-\rvX_3$ forms a Markov chain. In other words, the actions of Agent 3 should be conditionally independent of the actions of Agent 1 given the actions of Agent 2. Note that the \jkl{\emph{communication}} topology (a line network here) \emph{matches}  this \jkl{\emph{coordination}} structure, since Agent 3 only communicates with Agent 1 through Agent 2. Specializing Corollary~\ref{cor:projection_region}, we obtain 
\begin{align}
  \calC^*(q_{\rvX_1\rvX_2\rvX_3})=\left\{
        (R_{12},R_{23}):      \begin{array}{l}
        R_{12}\geq\avgI{\rvX_2;\rvX_1}\\
        R_{23}\geq\avgI{\rvX_3;\rvX_1}
      \end{array}
\right\}.
\end{align}
Let us consider a modified communication topology as shown in Fig.~\ref{fig:modified_line}, such that the roles of Agent 3 and Agent 2 are swapped. There is now a \emph{mismatch} between the communication topology and the desired behavior, since Agent 3 becomes the communication bottleneck. This effect is captured by specializing again Corollary~\ref{cor:projection_region}, which yields the region
\begin{align}
  \calC^*(q_{\rvX_1\rvX_2\rvX_3}) = \left\{
        (R_{13},R_{32}):      \begin{array}{l}
        R_{13}\geq\avgI{\rvX_2;\rvX_1}\\
        R_{32}\geq\avgI{\rvX_2;\rvX_1}
      \end{array}
\right\}.
\end{align}
Since $\avgI{\rvX_2;\rvX_1}\geq \avgI{\rvX_3;\rvX_1}$ by the data-processing inequality, we see that the mismatch translates into a penalty in terms of the communication rates between agents.
\section{Achievability Proof}
\label{sec:achievability-proof}

Because of space constraints, we do not detail some of the more technical steps of the proof. Let $\rvU, \rvV, \rvW, \rvX_1,\rvX_2,\rvX_3$ be discrete random variables with joint distribution
\begin{multline*}
  p(u,v,w,x_1,x_2,x_3)\eqdef  W(x_1|u,w) W(x_2|u,v,w) \\
  W(x_3|v,w)p(u|w)p(v|w)p(w)
\end{multline*}
such that the marginal $p_{\rvX_1\rvX_2\rvX_3}$ satisfies $p(x_1,x_2,x_3) = q(x_1,x_2,x_3)$. Note that at least one such distribution exists (with $\rvU\eqdef\rvX_2$, $\rvW\eqdef \rvX_3$, $\rvV\eqdef \rvX_3$) and that it belongs to the set $\calS_{\text{in}}$.

\begin{figure}[b]
  \centering
    \scalebox{0.7}{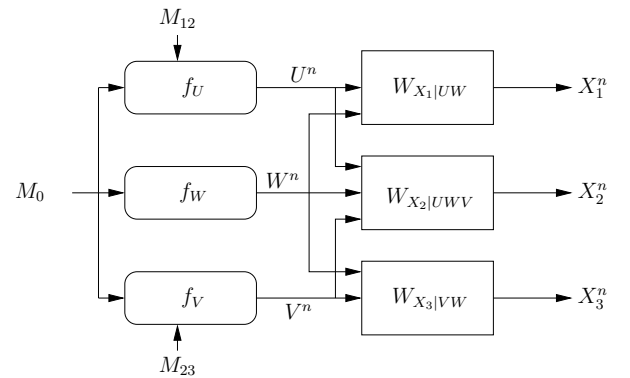}
  \caption{Intermediate problem used in achievability proof.}
  \label{fig:intermediate}
\end{figure}
Following the idea in~\cite{Cuff2008}, we solve the coordination problem by constructing a code for an intermediate problem, which is illustrated in Fig.~\ref{fig:intermediate}. Three uniformly distributed messages $M_0\in\intseq{1}{2^{nR_0}}$, $M_{12}\in\intseq{1}{2^{nR_{12}}}$ and $M_{23}\in\intseq{1}{2^{nR_{23}}}$ are encoded into codewords $\rvU^n\in\calU^n$, $\rvV^n\in\calV^n$ and $\rvW^n\in\calW^n$ using the following encoding functions:
\begin{itemize}
\item $f_{\rvU}:\intseq{1}{2^{nR_0}}\times\intseq{1}{2^{nR_{12}}}\rightarrow \calU^n$;
\item $f_{\rvW}:\intseq{1}{2^{nR_{0}}}\rightarrow \calW^n$;
\item $f_{\rvV}:\intseq{1}{2^{nR_0}}\times\intseq{1}{2^{nR_{23}}}\rightarrow \calV^n$.
\end{itemize}
The codewords $\rvU^n$, $\rvW^n$, $\rvV^n$, are then transmitted into channels with transition probabilities $W_{\rvX_1|\rvU\rvW}$, $W_{\rvX_2|\rvU\rvV\rvW}$, $W_{\rvX_3|\rvV\rvW}$, respectively. The encoders induce a joint distribution between messages, and channel outputs, which we denote by $\hat{p}_{\rvX_1^n\rvX_2^n\rvX_3^n\rvM_0\rvM_{12}\rvM_{23}}$.
The problem is to identify triplets $(R_0,R_{12}, R_{23})$ that are achievable, in the sense that there exists a sequence of encoders with increasing length $n$ such that
\begin{align}
  &\lim_{n\rightarrow\infty}\V{\hat{p}_{\rvX_1^n,\rvX_2^n,\rvX_3^n},q_{\rvX_1^n\rvX_2^n\rvX_3^n}}=0\label{eq:resolvability}\\
  \text{and }&\lim_{n\rightarrow\infty}\V{\hat{p}_{\rvX_1^n,M_0},\hat{p}_{\rvX_1^n}\hat{p}_{M_0}}=0\label{eq:secrecy}
\end{align}
The constraint in Eq.~\eqref{eq:resolvability} is a \emph{channel
  resolvability} constraint, which requires the encoders to simulate the
distribution $q_{\rvX_1^n\rvX_2^n\rvX_3^n}$, while the constraint in
Eq.~\eqref{eq:secrecy} is a \emph{secrecy} constraint, which requires the
message $\rvM_0$ to be independent of the output $\rvX_1^n$. \jkl{This ensures the compatibility of a code for the intermediate problem in Fig.~\ref{fig:intermediate} with a code for the original problem in~Fig.~\ref{fig:coordination_setting}.}

Since all channels considered here have multiple inputs, the key conceptual tool we rely on is a variation of multiple-access channel resolvability~\cite{Steinberg1998,Yassae2010,Pierrot2011a}.

To simplify notation in the sequel, we denote $Z^n\eqdef \rvX_1^n\rvX_2^n\rvX_3^n$, which we use when the analysis does not require us to treat $\rvX_1^n$, $\rvX_2^n$, and $\rvX_3^n$ separately.

We start by generating three codebooks randomly.
\begin{itemize}
\item We generate $2^{nR_0}$ sequences, labeled $w^n_i=(w_{i,1},\cdots,w_{i,n})$, independently according to $\prod_{\ell=1}^np_{W}(w_{i,\ell})$;
\item For each $w^n_i$, we generate $2^{R_{12}}$ sequences, labeled $u^{n}_{ij}$, independently according to  $\prod_{\ell=1}^np_{U|W}(u_{ij,\ell}|w_{i,\ell})$;
\item For each $w^n_i$, we generate $2^{R_{23}}$ sequences, labeled $v^{n}_{jk}$, independently according to  $\prod_{i=1}^np_{V|W}(v_{jk,\ell}|w_{i,\ell})$.
\end{itemize}
The indices of sequences in the codewords define the encoding functions $f_\rvU$, $f_\rvW$, $f_\rvV$ of the code.

Next, we analyze $\E{\V{\hat{P}_{\rvZ^n},q_{\rvZ^n}}}$, where the
expectation is over the randomly generated code and $\hat{P}$ denotes the
probability $\hat{p}$ for a random code. We denote by
$\sts{\delta}{n}{\rvZ}$ the $\delta$-typical set for the \jkl{distribution}
$p_{\rvZ}$ and by $\sts{\delta}{n}{\rvU\rvV\rvW\rvZ|z^n}$ the
$\delta$-conditional typical set for the \jkl{distribution}  $p_{\rvU\rvV\rvW\rvZ}$ and $z^n\in\sts{\delta}{n}{\rvZ}$.  Upon defining 
\begin{multline*}
  \hat{p}^{(1)}_{\rvZ^n}(z^n)\eqdef \sum_{i,j,k}W(z^n|u^n_{ij},v^n_{ik},w^n_i)2^{-n(R_0+R_{12}+R_{23})}\\
  \mathbf{1}\left\{(u^n_{ij},v^n_{ik},w^n_i)\in\sts{\delta}{n}{\rvU\rvV\rvW\rvZ|z^n}\right\}
\end{multline*}
and
\begin{multline*}
  \hat{p}^{(2)}_{\rvZ^n}(z^n)\eqdef \sum_{i,j,k}W(z^n|u^n_{ij},v^n_{ik},w^n_i) 2^{-n(R_0+R_{12}+R_{23})}\\
  \mathbf{1}\left\{(u^n_{ij},v^n_{ik},w^n_i)\notin\sts{\delta}{n}{\rvU\rvV\rvW\rvZ|z^n}\right\},
\end{multline*}
so that $\hat{p}_{Z^n}(z^n)=\hat{p}^{(1)}_{\rvZ^n}(z^n)+\hat{p}^{(2)}_{\rvZ^n}(z^n)$, and using the triangle inequality repeatedly, one obtains the following upper bound
\begin{align*}
\V{\hat{P}_{\rvZ^n},q_{\rvZ^n}}&\leq \sum_{z^n\in\sts{\delta}{n}{\rvZ}}\abs{\hat{P}^{(1)}_{\rvZ^n}(z^n)-\E{\hat{P}^{(1)}_{\rvZ^n}(z^n)}}\\
 &\phantom{--}+\sum_{z^n\in\sts{\delta}{n}{\rvZ}}\abs{\hat{P}^{(2)}_{\rvZ^n}(z^n)-\E{\hat{P}^{(2)}_{\rvZ^n}(z^n)}}\\
    &\phantom{--}+    \sum_{z^n\notin\sts{\delta}{n}{\rvZ}}\abs{\hat{P}_{\rvZ^n}(z^n)-q_{\rvZ^n}(z^n)}.
  \end{align*}
  One can then show that
\begin{align*}
  &\lim_{n\rightarrow\infty}\E{\sum_{z^n\notin\sts{\delta}{n}{\rvZ}}\abs{\hat{P}_{\rvZ^n}(z^n)-q_{\rvZ^n}(z^n)}}=0,\\
  &\lim_{n\rightarrow\infty}\E{\sum_{z^n\in\sts{\delta}{n}{\rvZ}}\abs{\hat{P}^{(2)}_{\rvZ^n}(z^n)-\E{\hat{P}^{(2)}_{\rvZ^n}(z^n)}}} =0.
\end{align*}
The last sum is upper bounded using Jensen's inequality as
\begin{multline}
  \E{\sum_{z^n\in\sts{\delta}{n}{\rvZ}}\abs{\hat{P}^{(1)}_{\rvZ^n}(z^n)-\E{\hat{P}^{(1)}_{\rvZ^n}(z^n)}}}\\
  \leq \sum_{z^n\in\sts{\delta}{n}{\rvZ}}\sqrt{\E{(\hat{P}^{(1)}_{\rvZ^n}(z^n))^2}-\E{\hat{P}^{(1)}_{\rvZ^n}(z^n)}^2}.\label{eq:bound_jensen}
\end{multline}
Note that $\E{(\hat{P}^{(1)}_{\rvZ^n}(z^n))^2}$ is written explicitly as
\begin{align*}
  \E{(\hat{P}^{(1)}_{\rvZ^n}(z^n))^2}= \sum_{ijk}\sum_{i'j'k'} A_{ijki'j'k'}\text{ with}
\end{align*}\vspace{-20pt}
\begin{multline}
A_{ijki'j'k'}\eqdef \mathbb{E}\left(W(z^n|U^n_{ij},V^n_{ik},W^n_i) W(z^n|U^n_{i'j'},V^n_{i'k'},W^n_{i'})\right.\\
  \mathbf{1}\left\{(U^n_{ij},V^n_{ik},W^n_i)\in\sts{\delta}{n}{\rvU\rvV\rvW\rvZ|z^n}\right\} \\
\left.  \mathbf{1}\left\{(U^n_{i'j'},V^n_{i'k'},W^n_{i'})\in\sts{\delta}{n}{\rvU\rvV\rvW\rvZ|z^n}\right\}\right)
\end{multline}
Because of the properties of the random code generation procedure, the analysis of the sum can be split into 5 parts.

If $i\neq i'$, and for any $j,k,j',k'$, we obtain
  \begin{align}
    \sum_{i,j,k,i'\neq i,j',k'}A_{ijki'j'k'} \leq \E{\hat{p}^{(1)}_{\rvZ^n}(z^n)}^2\label{eq:bound_aijk_1}
  \end{align}
If $i=i'$, $j=j'$, $k=k'$, we obtain
  \begin{align}
    \sum_{i,j,k}A_{ijkijk} \leq  2^{-n(R_0+R_{12}+R_{23}+\avgH{\rvZ|\rvU\rvV\rvW}+\avgH{\rvZ}-\calO(\delta))}\label{eq:bound_aijk_2}
  \end{align}
If $i=i'$, $j=j'$, $k\neq k'$, we obtain
  \begin{align}
    \sum_{i,j,k,k'\neq k}A_{ijkijk'} \leq  2^{-n(R_0+R_{12}+\avgH{\rvZ|\rvU\rvW}+\avgH{\rvZ}-\calO(\delta))}\label{eq:bound_aijk_3}
  \end{align}
If $i=i'$, $j\neq j'$, $k=k'$, we obtain
  \begin{align}
    \sum_{i,j,k,j'\neq j}A_{ijkij'k} \leq  2^{-n(R_0+R_{23}+\avgH{\rvZ|\rvV\rvW}+\avgH{\rvZ}-\calO(\delta))}\label{eq:bound_aijk_4}
  \end{align}
If $i=i'$, $j\neq j'$, $k\neq k'$, we obtain
  \begin{align}
    \sum_{i,j,k,j'\neq j,k'\neq k}A_{ijkij'k'} \leq  2^{-n(R_0+\avgH{\rvZ|\rvW}+\avgH{\rvZ}-\calO(\delta))}\label{eq:bound_aijk_5}
  \end{align}
Substituting Eq.~\eqref{eq:bound_aijk_1}-\eqref{eq:bound_aijk_5} into Eq.~\eqref{eq:bound_jensen} and using the bound $\card{\sts{\delta}{n}{\rvZ}}\leq 2^{n(\avgH{\rvZ}+\calO(\delta))}$, one can finally show that if
\begin{align}
  \begin{split}
    R_0+R_{12}+R_{23} &\geq \avgI{\rvU\rvV\rvW;\rvZ},\\
    R_0+R_{12} &\geq \avgI{\rvU\rvW;\rvZ},\\
    R_0+R_{23} &\geq \avgI{\rvV\rvW;\rvZ},\\
    R_0&\geq \avgI{\rvW;\rvZ},
  \end{split}  \label{eq:bound_1}
\end{align}
then
\begin{align*} 
\lim_{n\rightarrow\infty}\E{\sum_{z^n\in\sts{\delta}{n}{\rvZ}}\abs{\hat{p}^{(1)}_{\rvZ^n}(z^n)-\E{\hat{p}^{(1)}_{\rvZ^n}(z^n)}}} =0.
\end{align*}
The analysis of $\E{\V{\hat{p}_{\rvX_1^n,M_0},\hat{p}_{\rvX_1^n}\hat{p}_{M_0}}}$ follows similar steps once using the triangle inequality repeatedly to show
\begin{align*}
  \V{\hat{p}_{\rvX_1^n,M_0},\hat{p}_{\rvX_1^n}\hat{p}_{M_0}} \leq 2 \E[M_0]{\V{\hat{p}_{\rvX_1^n|M_0},q_{\rvX_1^n}}}.
\end{align*}
The secrecy constraint then reduces to another multiple-access channel resolvability constraint, and we can show that if
\begin{align}
  \begin{split}
    R_{12}+R_{23}&\geq \avgI{\rvU\rvV\rvW;\rvX_1},\\
    R_{12}&\geq \avgI{\rvU\rvW;\rvX_1},\\
    R_{23}&\geq \avgI{\rvV\rvW;\rvX_1},
  \end{split}\label{eq:bound_2}
\end{align}
then $\lim_{n\rightarrow\infty}\E{\V{\hat{p}_{\rvX_1^n,M_0},\hat{p}_{\rvX_1^n}\hat{p}_{M_0}}} =0$. Using Markov's inequality, one then shows the existence of a sequence of codes achieving the rates $(R_0,R_{12},R_{23})$ satisfying the constraints in Eq.~\eqref{eq:bound_1} and Eq.~\eqref{eq:bound_2}. 

All that remains to show now is how the code for the intermediate problem can be used as a code for the original coordination problem. We choose the encoding and decoding for the coordination problem to operate as follows.
\begin{itemize}
\item Agent 1 generates message $m_{12}$ from his actions $x_1^n$ and the common randomness $m_0$ according to $\hat{p}(m_{12}|x_1^n,m_0)$; Agent 1 then sends $m_{12}$ to Agent 2;
\item Agent 2 generates a message $m_{23}$ uniformly at random and receives $m_{12},m_0$; Agent 2 then simulates the transmission of the codewords $u^n_{m_0,m_{12}}$, $v^n_{m_0,m_{23}}$, $w^n_{m_0}$ through the channel with transition probabilities $W_{\rvX_2|\rvU\rvV\rvW}$ to obtain his actions $x_2^n$; Agent 2 then sends $m_{23}$ to Agent 3;
\item Agent 3 receives $m_{23},m_{0}$ and simulates the transmission of the codewords $v^n_{m_0,m_{23}}$ and $w^n_{m_0}$ through the channel with transition probabilities $W_{\rvX_3|\rvV\rvW}$ to obtain his actions $x_3^n$.
\end{itemize}
We denote the joint distribution induced by this scheme by $\tilde{p}_{\rvX_1^n\rvX_2^n\rvX_3^n\rvM_0\rvM_{12}\rvM_{23}}$. Using the triangle inequality,\vspace{-5pt}
\begin{multline*}
  \V{\tilde{p}_{\rvX_1^n\rvX_2^n\rvX_3^n},q_{\rvX_1^n\rvX_2^n\rvX_3^n}} \leq   \V{\tilde{p}_{\rvX_1^n\rvX_2^n\rvX_3^n},\hat{p}_{\rvX_1^n\rvX_2^n\rvX_3^n}}\\
  +   \V{\hat{p}_{\rvX_1^n\rvX_2^n\rvX_3^n},q_{\rvX_1^n\rvX_2^n\rvX_3^n}}.
\end{multline*}
The first term on the right-hand side vanishes since the code satisfies Eq.~\eqref{eq:secrecy} \jkl{after some intermediate steps}, while the second term  vanishes since the code satisfies Eq.~\eqref{eq:resolvability}. Hence, the achievable coordination rates are exactly the achievable rates for the intermediate problem. Combining all rate constraints and recalling the construction of $\rvU\rvV\rvW$, one obtains that $\bigcup_{p\in\calS_{\text{in}}}\calR(p)\subset \calC$.

\vspace{-1ex}
\section{Converse Proof}
\label{sec:converse-proof}
Let $(R_0,R_{12},R_{23})$ be achievable coordination rates, so that for all
$\epsilon>0$, there exists a $(2^{nR_0},2^{nR_{12}},2^{nR_{23}},n)$ line
coordination code for which
$\V{p_{X_1^nX_2^nX_3^n},q_{X_1^n,X_2^n,X_3^n}}\leq \epsilon$. For brevity,
we only sketch the calculation of some converse constraints. \pagebreak[2]
Following the steps as in~\cite{Cuff2008}, we obtain
\[
  R_0+R_{12}+R_{23}
\geq   \avgI{\rvX_{1,\rvQ}\rvX_{2,\rvQ}\rvX_{3,\rvQ};\rvM_0\rvM_{12}\rvM_{23}\rvQ}-\delta(\epsilon),
\]
where the random variable $\rvQ$ is uniformly distributed in $\intseq{1}{n}$ and independent of all others, and $\delta(\epsilon)$ denotes a function such that $\lim_{\epsilon\rightarrow0}\delta(\epsilon)=0$. Upon introducing $X_i\eqdef X_{i,\rvQ}$ for $i\in\{1,2,3\}$, $\rvU\eqdef\rvM_{12}$, $\rvV\eqdef\rvM_{23}$ and $\rvW\eqdef \rvM_0$, we obtain
\begin{align*}
  R_0+R_{12}+R_{23}\geq \avgI{\rvX_1\rvX_2\rvX_3;\rvU\rvV\rvW}-\delta(\epsilon).
\end{align*}
Similar bounds are obtained for $R_0+R_{12}$, $R_0+R_{23}$ and $R_0$. Next, using the independence of $\rvM_0$ and $\rvX_1^n$, we can also show \vspace{-10pt}
\begin{align*}
  R_{12}+R_{23}\geq \avgI{\rvX_1;\rvU\rvV\rvW}.
\end{align*}
Similar bounds are obtained for $R_{12}$ and $R_{23}$. Finally, using a functional dependence graph, one can check that the following Markov chains hold:
\begin{align*}
  \rvX_1-\rvU\rvW-\rvV\rvX_2\rvX_3\text{ and }
  \rvX_1\rvX_2\rvU-\rvV\rvW-\rvX_3.
\end{align*}
\jkl{Note that the constraint
$\rvU-\rvW-\rvV$ from  the achievability
proof does \emph{not} hold here, which puts an additional restriction  
on the coding scheme used in the achievability proof and shows that this
scheme is sub-optimal in general.} Combining all constraints and taking the limit as $\epsilon$ goes to zero yields the desired converse (see~\cite[Lemma VI]{Cuff2012} for a careful justification).

\vspace{-5pt}
\section{Acknowledgement}
The research was supported in part by NSF grants CCF-0830666 and CCF-1017632, and the CNRS grant PEPS PhySecNet.

\bibliographystyle{IEEEtran}
\bibliography{isit2013}

\end{document}